\documentstyle[aps,pre,psfig,twocolumn]{revtex}
\begin{document}
\newcommand{\avg}[1]{\langle{#1}\rangle}
\newcommand{\req}[1]{(\ref{#1})}
\def\be{\begin{equation}}
\def\ee{\end{equation}}
\def\bc{\begin{center}}
\def\ec{\end{center}}
\def\bea{\begin{eqnarray}}
\def\eea{\end{eqnarray}}
\title{Scale Invariant Dynamics of Surface Growth}
\author{C. Castellano$^{(1)}$, M. Marsili$^{(3)}$,
M. A.  Mu{\~{n}}oz$^{(1,2)}$, 
and L. Pietronero$^{(1,2)}$}
\address{$^{(1)}$ The Abdus Salam International 
Centre for Theoretical Physics, P. O. Box 586,
I-34100 Trieste, Italy\\
$^{(2)}$ Dipartimento di Fisica and Unit\`a INFM,
Universit\`a di Roma ``La Sapienza'', I-00185 Roma, Italy\\
$^{(3)}$
International School for Advanced Studies
(SISSA) and Unit\`a INFM, via Beirut 2-4, Trieste I-34014, Italy
}
\maketitle 

\begin{abstract}
We describe in detail and extend a recently introduced nonperturbative
renormalization group (RG) method for surface growth.
The scale invariant dynamics which is the key ingredient 
of the calculation is obtained 
as the fixed point of a RG transformation relating the
representation of the microscopic process at two different 
coarse-grained scales.
We review the RG calculation for systems in the Kardar-Parisi-Zhang
universality class and compute the roughness exponent for the strong coupling
phase in dimensions from $1$ to $9$.
Discussions of the approximations involved
 and possible improvements are also presented.
Moreover, very strong evidence of the
absence of a finite upper critical dimension for KPZ growth
is presented. Finally, we apply the method to the linear
Edwards-Wilkinson dynamics where we reproduce the known exact
results, proving the ability of the method to capture qualitatively
different behaviors.
\end{abstract}

\narrowtext
\section{Introduction}

The idea that scale invariance is at the origin of
critical phenomena associated with equilibrium  second order phase
transitions has proven to be very fruitful. The analysis of scale
transformations in equilibrium statistical systems, now known as
renormalization group (RG), has indeed allowed for the explicit calculation
of critical exponents and, moreover, has led to the introduction of new
fundamental concepts such as scaling and universality.

The extension of the RG approach to non-equilibrium
phenomena, where scale invariance is widely observed, and
the identification of new universality classes, is of great 
importance from both theoretical and practical points of view. 
Technically, the RG ideas can be implemented in different
ways. The most standard one for systems at 
equilibrium is to consider their stationary
probability distribution written in terms
of continuum coarse-grained fields, and study them
perturbatively around their corresponding upper critical dimension. 
The most systematic
way to extend the previous methods to non-equilibrium systems,
where in general the stationary probability distribution
is not known, is to cast them into a continuum
 dynamical equation \cite{HH},
 or equivalently into a generating functional 
or action \cite{ZJ}. This last one can,
 in principle, be treated using the
same perturbative techniques developed to deal 
with equilibrium systems.
 However, there are some cases
where perturbative methods around a mean field 
solution are not suitable. In these cases the
$\epsilon$-expansion fails to give information on the
relevant physics. This turns out to be
governed by a strong coupling,  
perturbatively inaccessible fixed point.
The prototypical example of this class of systems
is the well known 
Kardar-Parisi-Zhang (KPZ) equation for surface growth \cite{KPZ},
where the properties of rough surfaces have not been
so far explained satisfactorily in generic spatial dimension.
This is a problem of great theoretical importance since the
KPZ describes not only the properties of rough
surfaces\cite{HZ,Krug,Laszlo}, but is also related to the Burgers
equation of turbulence \cite{Burger},
to directed polymers in random media \cite{DiP}, 
and to systems with multiplicative noise \cite{MN}. 
In particular, one of the most debated issues in this context is the
existence of an upper critical dimension, above which the system
is well described by the nontrivial infinite dimensional limit\cite{dinfi}.

Although the usual approach fails for KPZ, the presence of generic scale
invariance suggests that also for this system the basic idea of the
RG approach should be applicable in some form.

Real space approaches have
proven useful wherever standard perturbative techniques fail
\cite{VL,FST,Binney}. This, for example, is the case of fractal 
growth, and in particular for Diffusion-Limited-Aggregation\cite{FST}.
However, the attempts to apply standard  real space techniques to the KPZ
problem (and to surface growth in general) fail because of a fundamental
technical difficulty: The anisotropy of the scaling properties of the system.
That is, in order to cover with blocks (in the Kadanoff sense \cite{Binney})
a surface, isotropic blocks cannot be used:
Lengths in different directions must scale in different ways,
and the relative shape of blocks has to depend upon the scale via an
exponent that is unknown.
This makes conceptually non-trivial the application of real space RG
procedures to surface growth processes.

In this paper we investigate the scale invariant properties
of generic interface growth processes
through the introduction of a real-space
method. To achieve this goal we introduce some new
ingredients permitting us to overcome the aforementioned problem.
In particular, we introduce the idea that the statistical 
properties of growing surfaces on large scales can be described
in terms of an effective scale invariant dynamics for renormalized blocks.
Such dynamics is the fixed point of a RG transformation relating the
parameters of the dynamics at different coarse-graining levels.
The study of the RG flow, of the fixed points and of their stability
gives the universality classes and their associated exponents.

As a first application of the method, we study the
KPZ growth dynamics and obtain accurate
 estimates for the roughness exponent (when compared with
numerical results) in spatial dimensions from $d=1$ to $d=9$.
Furthermore, an analytical approximation allows us to exclude
the existence of a finite upper critical dimension for KPZ dynamics
and suggests that the roughness exponent decays as $1/d$ for large
dimensions, shedding light on a currently much debated issue.

In order to show the generality of the new real space scheme
and test its accuracy, we also apply it to the well known
linear theory, the Edwards-Wilkinson (EW) equation.  
We reproduce the expected behavior in different dimensions,
confirming the general applicability of the method.

The paper is organized as follows. In section II we present
the general RG method, the main concepts, the basic equations,
and discuss all the approximations involved.
In section III we review some results associated with KPZ growth
and apply the new RG method to such problem. We present some 
simple analytical approximations, explicit results for spatial
dimensions up to $d=9$, and discuss the
large dimensional limit in detail. In section IV we report
results on the analysis of the Edwards-Wilkinson equation. 
In section V a critical discussion of the method and of the results is
reported.
Partial accounts of the work presented here have already been
published recently, with a slightly different
notation\cite{Castellano98a,Castellano98b}. 

\section{Real Space RG for Surface Growth}

In order to present the RG method let us
consider a generic surface growth model where the height
is a single-valued function  $h(\vec x,t)$, with $\vec x$ the position in
a $d$-dimensional substrate and $t$ denoting time. The possibility
of having overhangs will not be considered here, as they are known
to be irrelevant for the asymptotic behavior of KPZ-like growth
\cite{overhangs}. 
The generic growth model under consideration 
can be either described at the microscopic level by a stochastic
equation or by a discrete dynamical rule.
In the first case $h$ and $\vec x$ are continuous variables, 
while in the latter they are discrete.

The roughness of a system, when considered on a substrate 
of linear size $L$, is defined by
\be
W^2(L,t)= {1 \over L^d} \sum_{\vec x} \left [h(\vec x,t)-
{\bar h}(t) \right]^2,
\ee
where
\be
{\bar h}(t) = {1 \over L^d} \sum_{\vec x} h(\vec x,t).
\ee
If we start the growth process from a flat configuration,
for short times the roughness grows as
\be
W(L,t) \sim t^\beta
\ee
until it reaches a stationary state characterized by
\be
W(L) \sim L^\alpha.
\ee
The crossover between the two behaviors occurs at a characteristic
 time $t_s$, that scales with $L$ as $L^z$.
This is the time scale over which correlations decay in the stationary state.
The exponents $\alpha$, $\beta$ and $z$ are to a large extent universal
for many different growth processes,
and are related by the trivial scaling relation $\beta=\alpha/z$.

We now introduce the real space renormalization group (RSRG)
procedure aimed at the study the stationary state and
in particular at the determination of the roughness exponent $\alpha$.
The following subsections
are structured as follows. In A) we introduce the geometric elements
or blocks 
(equivalent to the Kadanoff blocks in standard RSRG methods)
suitable to deal with anisotropic situations. In B) we  discuss
the effective dynamics of the previously defined blocks at a generic scale. 
In C) we introduce the RG equation and explain how the roughness
exponent is determined. Finally in D) we analyze critically
the approximations involved in general in the method.

\subsection{Geometric description}

The first nontrivial problem in the development of a RSRG approach
is to find a sensible description of the geometry of the growing surface
at a generic scale, i. e. how to build the analog of a block-spin
transformation\cite{Binney}.
Given the anisotropy of the system, the shape of the blocks must depend
on the scale. Therefore, subdividing a cell in subcells is not a
feasible task and the explicit construction of the block-spin transformation
is not possible.

\begin{figure}
\centerline{\psfig{figure=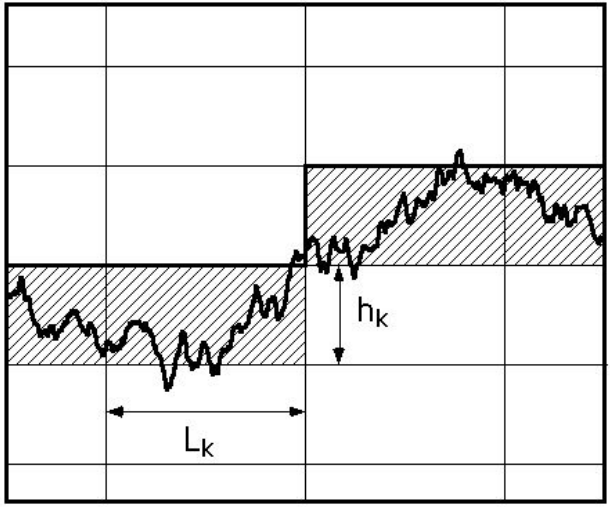,width=10cm}}
\caption{Covering procedure of a particular microscopic surface with
cells of size $L_k \times h_k$. Blocks below (above)
the surface are considered to be occupied (empty).}
\label{Fig01}
\end{figure}

Hence we develop an alternative strategy.
To obtain a description at a generic scale $k$ of the growing surface,
we consider a partitioning of the $(d+1)$-dimensional space in cells
of lateral size $L_k = L_0 b^k$ and vertical size $h_k$.
Here $b$ is a constant and $k$ labels the scale (Fig.~\ref{Fig01}).

A cell is declared to be empty or filled according to a majority rule.
In this way we pass from the microscopic description $h(\vec x,t)$ to a
coarse-grained one at scale $k$, fully defined by the number
$h(i,k,t)$ of filled blocks in the column $i$.
Heights at scale $k$ are measured in units of $h_k$. 
The only characteristic vertical length at scale $k$ is that 
fixed by typical intrinsic fluctuations of the surface of a lateral size
$L_k$. This suggests to take
\be
h_k = \sqrt{c} W(L_k) \sim L_k^\alpha,
\label{h_k}
\ee
where $\sqrt{c}$ is a proportionality constant that will be discussed later.
This equation expresses the requirement of scale invariance in 
the geometric description. Any other choice would result either
in a redundant description (if $h_k/W(L_k)\to 0$ as $k\to\infty$)
where too many (infinite) blocks would be needed to describe fluctuations
in the same column, or in a too coarse description (if $h_k/W(L_k)\to 
\infty$ as $k\to\infty$). By imposing Eq. (\ref{h_k}),
we always have a meaningful covering of the surface
upon scale changes.  Observe that since in general $\alpha \neq 1$,
the shape (i.e. the ratio of vertical to horizontal length) of the
blocks changes with the scale $k$.
Contrarily to the usual RG approach, the definition of the
block-spin transformation depends explicitly on the roughness
exponent $\alpha$, the calculation of which
is the final goal of the method.

The constant $c$ in Eq.~(\ref{h_k}) fixes the unit of measure of
our blocks. Its optimal value can be determined as follows.
The distribution of microscopic height fluctuations within a block
can be mapped into an {\em effective} distribution
with the same average $\bar h$ and standard deviation.
For simplicity we take it to be bimodal
\begin{eqnarray}
P[h(x)]  =  p &\delta & \{h(x)-[{\bar h}+(1-p) h_k]\} \nonumber \\  + 
(1-p) &\delta & \{h(x)-[{\bar h}-p h_k]\}.
\label{pofh}
\end{eqnarray}
This distribution results from mapping all points with microscopic
height larger than $\bar h$ to $\bar h+ (1-p) h_k$
and those smaller than $\bar h$ to $\bar h -p h_k$.
The parameter $p$ describes the degree of asymmetry of the
distribution: The fluctuations inside a block
can then be calculated, using Eq.~(\ref{pofh}), as
\be
W^2(L_k)= p (1-p) h^2_k,
\label{prop}
\ee
which implies that the constant $c$ is given by
\be
c = {1 \over p(1-p)}.
\ee
For a symmetric distribution $p=1/2$ and therefore $c=4$.
In general the height distribution is not symmetric, i. e.
there is some nonvanishing skewness and one must consider $c \ne 4$.

\subsection{Dynamic description}

The second step in the construction of the RG procedure is the
definition of the effective dynamics at a generic scale $k$, 
i.e. the determination of the growth rules for the blocks defined
in the previous subsection.
The effective dynamics will depend on a set of scale dependent parameters.
The changing of scale induces a flow in the parameter space
whose fixed points correspond to the scale invariant dynamics.

Analogously to what happens in the usual
application of the RG approach to equilibrium systems,
it may happen that mechanisms not appearing 
in the microscopic rule are generated upon coarse-graining.
In the language of equilibrium systems this means that operators not
included in the bare Hamiltonian can be generated iteratively.
Conversely, microscopic ingredients can prove to be irrelevant 
and be progressively eliminated when going to coarser scales.
Therefore, exactly as in the equilibrium case the choice of the
parametrization of the effective dynamics is not trivial:
Principles as the preservation
of symmetries and conservation laws must be the guidelines.
In general, the effective dynamics will be defined in terms of the
transition rates for the addition of occupied blocks
at a generic coarse-grained scale, that is,
\be
r[h(i,k) \to h'(i,k)] = r(x_k^1,x_k^2,\ldots,x_k^n).
\ee
The number of parameters $x_k^i$ is in principle arbitrary, although in
the applications presented below it will be limited to one \cite{bi}.
It is clear that the more complete the parametrization the better 
the final description of the statistical scale invariant state.
We will discuss this problem in detail in subsection D.

\subsection{The RG Equations}

So far we have defined the geometrical and dynamical aspects of the
coarse-graining procedure. These give us the necessary ingredients
to introduce the RG transformation.
The explicit derivation of it
is based on the following property of the roughness $W$.
Let us consider a $d$-dimensional 
system of linear size $L$ and partition it in $(L/b)^d$ blocks
of size $b^d$ (labeled by the index $j$).
It is straightforward to verify
that the total roughness can be decomposed as
\begin{eqnarray}
W^2(L) &= & {1 \over (L/b)^d} \sum_{j=1}^{(L/b)^d} \left \{ {1\over b^d}
\sum_{i \in j} \left[h(i)-{\bar h}(j) \right]^2 \right \}
 \nonumber \\ &+&
{1\over (L/b)^d}
\sum_{j=1}^{(L/b)^d} \left[{\bar h}(j)-{\bar h} \right]^2,
\end{eqnarray}
where ${\bar h}(j)$ is the average height within block $j$.
The interpretation of this formula is simple:
The first term on the right hand side is the
averaged value of the roughness within blocks of size
$b^d$, while  the second term is the fluctuation 
of the average value of $h$ among blocks.

In our coarse-graining procedure this property is read as follows:
If one takes $L=L_{k+1}=b L_k$ the first term on the right hand side
is $W^2(L_k)$, the total roughness within a block of size $L_k$; the second
is the roughness of the configuration in which blocks of size $L_k$ are
considered as flat objects.
This second contribution is obviously proportional to the square of the
height of a block $h_k^2$.
Hence, employing Eq. (\ref{h_k}),
\begin{eqnarray}
W^2(L_{k+1}) &= & W^2(L_k)+ \omega^2(b,k) h_k^2 \nonumber \\
& = &  \left[ 1 +
c\omega^2(b,k)\right] W^2(L_k)  \nonumber \\
 & = & F_b(k) W^2(L_k),
\label{main}
\end{eqnarray}
where  $\omega^2(b,k)$ is the roughness in the stationary state
of a system of $b^d$ sites of unit height that evolves according to
the dynamical rules specified by $(x_k^1,x_k^2,\ldots,x_k^n)$, and
\be
F_b(k) \equiv \left[ 1 + c\omega^2(b,k)\right].
\label{F}
\ee
Note that the  dependence on the scale $k$ is only through the
parameters $\{x_k\}$.

{\it Eq. (\ref{main}) is the equation that relates the width at scales $k$
and $k+1$.} In order to proceed further, we must evaluate 
the function $F_b(k)$, or equivalently  $\omega^2(b,k)$.
To do so, we identify all the possible surface configurations 
of a system of composed of $b^d$ sites, and write down a 
master equation for their associated probabilities $\rho_i$
\be
\partial_t \rho_i = \sum_j \rho_j P_{j \to i} - \rho_i \sum_j P_{i \to j}.
\ee
$P_{i \to j}$ is the rate for the transition between configuration $i$ and $j$
and depends on the set of parameters $\{x_k\}$.
Imposing the stationarity condition $\partial_t \rho_i=0$ and the
normalization $\sum_i \rho_i=1$ the master equation can be solved.
If we call $W^2_i$ the roughness of configuration $i$, then we can
write
\be
\omega^2(b,k) = \sum_i  \rho_i(k) W^2_i.
\ee
Depending on the particular structure of the master equation the explicit
solution of the previous equation may be difficult or impossible.
In such cases it may be more useful to determine $\omega^2(b,k)$ numerically
by performing (relatively small) Monte Carlo simulations.
We will describe examples of both analytical and numerical
computations of $\omega^2(b,k)$.

Let us suppose now that $\omega^2(b,k)$ has been determined.
Eq.~(\ref{main}) gives an explicit relation between the roughness
at two different scales.
Observe that so far the scale invariance idea has not been implemented.
We have just studied how the width changes upon changing the level
of description. 
The last task to be performed is the determination of the RG transformation
relating the parameters of the dynamics at scale $k$ with those at scale
$k+1$.
This is done by means of a self-consistency requirement for the description
of the same system at two different levels of detail, i.e. the total width
of a system should be independent on the size of the blocks we use
to describe it. To make this idea more precise, 
let us consider the case of a dynamics parametrized by only
one parameter $x_k$.
Let us take a system of size $L=L_{k+2}$. By applying Eq.~(\ref{main})
we have
\be
W^2(L_{k+2}) = F_b(x_{k+1}) W^2(L_{k+1}).
\ee
This procedure can be iterated again on each of the resulting 
systems of size $L_{k+1}$,
obtaining
\be
W^2(L_{k+2}) = F_b(x_{k+1}) F_b(x_k) W^2(L_k).
\ee
The same quantity can alternatively be computed by considering directly
the whole system as composed by $b^{2d}$ systems of size $L_k$.
Applying again Eq.~(\ref{main})
\be
W^2(L_{k+2}) = F_{b^2}(x_k) W^2(L_k).
\ee
Imposing the consistency of the two procedures one has an implicit
RG transformation for $x_k$
\be
F_b(x_{k+1}) = {F_{b^2}(x_k) \over F_b(x_k)},
\label{RGeq}
\ee
or explicitly
\be
x_{k+1} = R(x_k) \equiv F_b^{-1} \left[{F_{b^2}(x_k) \over F_b(x_k)} \right].
\ee
This equation provides the evolution of the parameter
under a change of scale.

If a fixed point $x^*$ such that
\be
x^*= R(x^*)
\label{FP}
\ee
exists, then the parameter $x^*$ characterizes
 the scale invariant dynamics
of the system.
The knowledge of it directly allows the determination of the exponent
$\alpha$. Since $W^2(L_{k+1})/W^2(L_k)$ is equal to $ b^{2 \alpha}$ we have
\begin{eqnarray}
\alpha & = & \lim_{k \to \infty} {1 \over 2} \log_b \left[
{W^2(L_{k+1}) \over W^2(L_k)} \right ] \nonumber \\
 & = & \lim_{k \to \infty} {1 \over 2}
\log_b F_b(k)={1 \over 2} \log_b F_b(x^*).
\label{alpha}
\end{eqnarray}
To analyze the stability of the
 fixed point we linearize the RG transformation
around it
\be
x_{k+1} - x^* = R(x_k) - x^* \simeq R'(x^*) (x_k-x^*).
\ee
Hence if $|R'(x^*)|<1$  the scale
 invariant dynamics specified by $x^*$
is an attractive fixed point under changes of scale.

Extension of the previous formalism to the case of $n$ parameters of the
dynamics is straightforward.
The $n$ RG transformations are obtained by imposing
the consistency of the description of the same system when divided in
$2^d$ and $4^d$ blocks, in $4^d$ and $16^d$ blocks, and so on \cite{bi}.

\subsection{Approximations}
 Let us discuss now the approximations involved in the method.
There are two steps where approximations come into play:
The first is the choice of the parametrization of the scale invariant
dynamics. The second is the computation of $\omega^2(b,k)$.

With respect to the first problem, it is reasonable to expect that
under coarse-graining the microscopic dynamics will flow towards a
scale invariant dynamics depending in principle on an infinite number
of parameters.
This proliferation is analogous to what happens in  RSRG
approaches to equilibrium systems.
The restriction to a finite (and small) number of parameters involves
unavoidably an approximation, due to the projection of the RG flow
onto the sub-space spanned by these parameters.

However, a very important  difference with respect to equilibrium 
critical phenomena is that here the scale invariant dynamics is
``self-organized critical'', that is, there are no relevant operators.
Only irrelevant fields, with negative scaling dimensions, need
to be parametrized.  The system is by definition {\em on the
critical manifold}
and, by iteration of the RSRG transformation, it converges to the 
stable fixed point, without any fine tuning of parameters.
 The projection onto a low-dimensional parameter 
space yields a projected RG flow which will share these same 
properties. The fixed point in this sub-space, being the 
projection of the actual fixed point in the high-dimensional 
space, will have the same qualitative properties. 
Even the simplest parametrization capturing the correct symmetries
of the dynamics can provide a quite accurate determination of the
properties of the system in this case.
On the contrary, when relevant fields are present, as in
second order phase transitions, truncation effects are quite 
dramatic. 
The reason is that relevant fields have, in general, 
a non-vanishing component on any discrete (lattice) operator.
 Any 
approximation due to truncation is amplified by the 
RG iteration thus driving the flow {\em out of} the critical 
manifold along the relevant directions. The determination of the 
fixed point becomes then very difficult\cite{Baillie}.

The second source of approximation  is the computation of $\omega^2(b,k)$.
As stated above this quantity  is the stationary roughness of a system 
composed of $b^d$ substrate sites evolving according to the dynamical rules
specified by the parameters $(x_k^1,\ldots,x_k^n)$.
This is a perfectly well defined 
quantity that may in principle be computed to
any degree of accuracy by solving the master equation.
However very often the structure  of the master equation is too complicated
to allow for a full solution.
One then has to devise suitable simplifications
to make the analytical computation feasible.
This involves approximations that affect the final result.
We will see an example of this way of proceeding and discuss how
the effect of the approximation can be controlled.
Alternatively, when $b$ and $d$ are not too large, one can resort
to the numerical evaluation of $\omega^2(b,k)$.
In practice this boils down to performing Monte Carlo simulations
of  small systems evolving with different values of the parameters
$\{x_k\}$.
It is important to stress that the MC procedure involves no approximation,
except for the fluctuations associated with statistical sampling.
We will describe below an example of this alternative way
of computing $\omega^2(b,k)$.

A delicate issue is also the choice of the boundary conditions.
In the conceptual framework described above, $\omega^2(b,k)$ is the
roughness of a section of size $b$ of an infinitely extended surface.
This would suggest the use of open boundary conditions. On the other
hand, when integrating out degrees of freedom relative to
height fluctuations inside the cell, one should not consider 
the fluctuation of the average slope. This slope effect is
eliminated if one uses closed (i.e. periodic)
boundary conditions. Even though the choice of the appropriate
boundary conditions is not trivial, 
we will see, in the KPZ case, that the use of periodic or open
boundary conditions has little effect on the value of the exponent.
Furthermore, one expects that both truncation errors and those
induced by neglecting fluctuations of boundary conditions
vanish as the parameter $b$ grows. 
Arguments in support of this conclusion are reported in the 
Appendix I.

\section{RG for KPZ dynamics}

\subsection{The problem of KPZ growth}

 The Kardar-Parisi-Zhang equation is the minimal continuum 
equation capturing the physics of rough surfaces. After 
appearing in 1986 \cite{KPZ}, an overwhelming number of studies has
been devoted to elucidate its properties\cite{HZ,Laszlo}.
It reads \cite{KPZ}
\be
{\partial h(x,t) \over \partial t} = \nu \nabla^2 h + {\lambda \over 2}
(\nabla h)^2 + \eta(x,t).
\ee
where $h(x,t)$ is a  height variable at time $t$ and position $x$ in a
$d$-dimensional
substrate of linear size $L$. $\nu$ and $\lambda$ are constants
 and $\eta$ is a gaussian white noise. As a consequence of 
a tilting (Galilean) invariance
\cite{Burger,HZ,Krug} $\alpha + z=2$, and since in general $z=\alpha/\beta$, 
there is only one independent exponent, say $\alpha$.
The difference between the KPZ equation and the linear equation
(Edwards-Wilkinson), describing surfaces growing under the effect of
random deposition and surface tension, is the presence of a nonlinear
term proportional to $\lambda$.  This nonlinear term is
generated by microscopical processes giving rise to lateral growth,
i.e. the fact that growth velocity is normal to the local surface
orientation.

Exact results\cite{HZ,Laszlo} indicate that in $d=1$ there is only
a rough phase for KPZ with $\alpha=1/2$.
Instead, standard field theoretical methods predict the presence of a
{\it roughening transition } above $d=2$ \cite{Terry}; i.e.,
there are two RG attractive
fixed points and an unstable fixed point separating them.
More specifically,  there is a gaussian fixed point
with $\alpha=0$ describing a flat phase (characterized by
a vanishing renormalized nonlinear coupling)
and a nontrivial one describing the rough phase (in which the
renormalized nonlinear coupling diverges in perturbation theory).
Perturbative methods fail to give any prediction
for the exponents in the rough phase. 
For $d > 2$, an $\epsilon$-expansion ($d=2+ \epsilon$) around the gaussian
solution can
be performed and the exponents at the roughening transition evaluated
to all orders in perturbation theory \cite{Nuclear,MN}. These
results seem to indicate the presence of an anomaly in
$d=4$ for the roughening transition. This has been
interpreted as an indication that $d_c=4$ is the upper critical dimension
for the rough phase, i.e., for the strong coupling fixed point
\cite{MCDC,Lassig,Bhatta}.
Above this dimension the exponents should take the values known
for $d=\infty$ \cite{dinfi}. 
Applications of non-perturbative methods such as functional renormalization
group\cite{hh} and Flory-type arguments \cite{Flory} also suggested
that $d_c=4$, in agreement with a $1/d$-expansion\cite{1/d}
around the $d=\infty$ limit.
The mode-coupling approximation led to contradictory results,
suggesting the existence of a finite
$d_c$ \cite{MCDC} or $d_c=\infty$ \cite{Yuhai}. Arguments
for a finite $d_c$ based on directed\cite{DP} or
invasion\cite{Cie} percolation have also been proposed.

On the other hand, numerical results seem to indicate that the exponent
$\alpha$ decays continuously with  
the system dimensionality up to $d=7$, excluding therefore 
$d=4$ as upper critical dimension \cite{debate}.

Finally, some doubts have been cast on the  validity of the continuum
approach to study rough surfaces  \cite{discre}.
Summing up, the issue of the behavior of the KPZ dynamics for $d \ge 2$
is a highly debated one, and it
is extremely desirable to have alternative  approaches shedding light
into the problem.
In what follows we present the application of our new RG scheme to KPZ growth.

\subsection{Simplest RG scheme}

\subsubsection{Parametrization of the dynamics}
\begin{figure}
\centerline{\psfig{figure=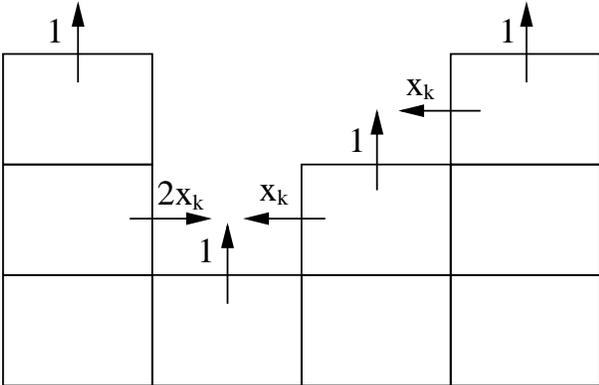,width=8cm,angle=-90}}
\caption{Growth rates for the KPZ dynamics in a typical configuration.}
\label{Fig02}
\end{figure}

The modelization of the dynamics at a generic scale should keep the number
of parameters to a minimum and catch all the relevant physical mechanisms
of the process.
The main feature of the KPZ dynamics is lateral growth.
Therefore we take as the only parameter defining the dynamics at a generic
scale $k$, the ratio $x_k$ between lateral and vertical growth
(i.e. random deposition).
More formally, the growth rate for the addition of an occupied block on
column $i$ is
\begin{eqnarray}
r_i & \equiv & r\left[h(i) \to h(i)+1 \right] \\
&= & 1 + x_k \sum_{j n.n. i} \max\left[0,h(j)-h(i)\right].
\label{r_i}
\end{eqnarray}
Eq.~(\ref{r_i}) states that the rate for lateral growth is
proportional to the difference in height between neighboring columns
(Fig.~\ref{Fig02}).
Overhangs, known to be irrelevant on large scales \cite{overhangs},
 are not allowed.
This dynamics can be seen as a generalization of the Eden growth model.

Few observations are in order.
We call the parameter $x$ appearing in Eq.~(\ref{r_i}) ``lateral growth''
parameter, but this is an abuse of language: $x_k$ cannot be identified 
with the parameter $\lambda$ of the KPZ equation.
Instead the term that multiplies
$x_k$ in Eq.~(\ref{r_i}) is a combination of the discretized Laplacian,
of the discretized square gradient and of other discrete operators.
The explicit dependence of
$x$ on $\nu$ and $\lambda$ cannot be disentangled.
Other parametrizations are clearly possible and they will be discussed below.

Eq.~(\ref{r_i}) has the nice feature that it contains as limiting cases
both the random deposition process ($x_k=0$) and the
infinitely strong ``lateral growth'' ($x_k=\infty$) leading to flat
surfaces. Most importantly,
it is easy to see that $x^*=\infty$ is, by construction, a fixed point 
of the RSRG with $\alpha=0$. This feature makes it possible the 
determination of the upper critical dimension above which the stable solution
leads to $\alpha=0$. 
In this situation we expect $x^*=\infty$ to be an attractive 
fixed point. 
Below the critical dimension, on the other hand, 
the fixed point $x^*=\infty$ must be
unstable and an intermediate fixed point with finite $\alpha$ must
appear. The RSRG accommodating for a fixed point at $x^*=\infty$,
naturally allows to address the issue of the upper critical dimensionality.

\subsubsection{d=1}
\begin{figure}
\centerline{\psfig{figure=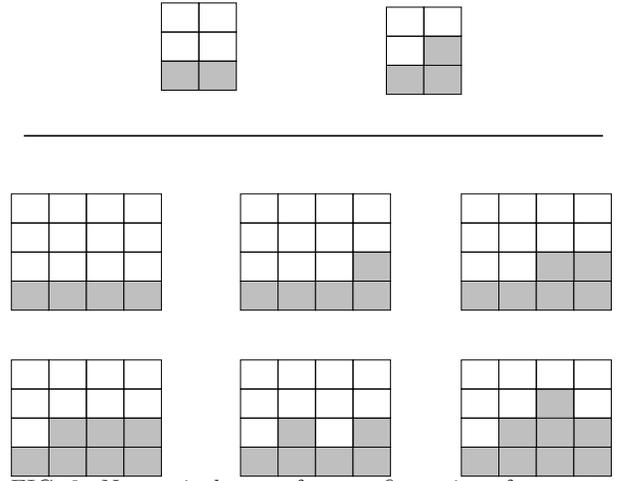,width=8cm,angle=-90}}
\caption{Nonequivalent surface configurations for a system of size 2 (top)
and 4 (bottom) with periodic boundary conditions and $\Delta h_{max}=1$.}
\label{Fig03}
\end{figure}

We restrict ourselves for the moment to the one-dimensional case and
illustrate in detail the application of the RG approach, i. e. the
computation of $\omega^2(b,k)$, the determination of the scale invariant
dynamics and of the exponent $\alpha$.
It is very instructive to consider first the dynamics Eq.~(\ref{r_i})
supplemented by the condition that the height difference between adjacent
columns is restricted to values such that 
($|\Delta h|\leq \Delta h_{max}$), with $\Delta h_{max}=1$.
This greatly reduces the number of possible surface configurations
allowing for a full analytical treatment.
For the system of size $b=2$, assuming periodic boundary conditions,
there are only 2 nonequivalent configurations,
while for the system of size 4 there are 6 of them (Fig.~\ref{Fig03}).
Using the definition Eq.~(\ref{r_i}) of the
growth rates for the addition of a block, one has simply, for $b=2$,
\begin{eqnarray}
P_{1 \to 1} &=& 0    \nonumber \\
P_{1 \to 2} &=& 2    \nonumber \\
P_{2 \to 1} &=& 1+2 x_k \nonumber \\
P_{2 \to 2} &=& 0.
\end{eqnarray}
In configuration 1 only vertical growth is possible (in two sites)
leading always to configuration 2.
Only one site can instead grow in configuration
 2 and the rate for this is the sum
of the rate of one vertical and two lateral contributions.
Hence the master equation reads
\bea
{\partial \rho_1 \over \partial t} &=& (1+2 x_k) \rho_2 - 2 \rho_1 \\
{\partial \rho_2 \over \partial t} &=& 2 \rho_1 -(1+2 x_k) \rho_2.
\eea
Imposing the stationarity condition $\partial_t \rho_1=\partial_t \rho_2=0$
and the normalization $\rho_1+\rho_2=1$ one has
\be
\rho_2={2 \over 3+2 x_k}
\ee
and then considering that the width associated with configurations
 1 and 2 is, respectively, $0$ and $1/4$
\be
\omega^2(2,x_k)={2 \over 4(3+2x_k)}.
\ee
For the system of size $b^2=4$ one finds in an analogous way
\be
\omega^2(4,x_k)={51+86 x_k +40 x_k^2 \over 4 (47+106 x_k+68 x_k^2 + 8 x_k^3)}.
\ee
Plugging these two expressions into the RG equation~(\ref{RGeq}) with
$c=4$  \footnote{In $d=1$ the distribution is known to be symmetric.}
one finds that the explicit form of the RG transformation is
\be
R(x)={293+804 x+636 x^2 +160 x^3 + 32 x^4 \over 2(59+148 x + 156 x^2
+ 64 x^3)}
\ee
and that there exists only one finite fixed point for
\be
x^* \simeq 2.08779 \ldots
\ee
Such a fixed point is attractive since
\be
R'(x^*)\simeq -0.03548\ldots
\ee
Hence no matter how small or large the microscopic value of $x$ is, 
upon coarse-graining the dynamics flows towards an attractive scale invariant
dynamics, characterized by a ratio $x^*$ of the lateral to vertical growth
rates.
The roughness associated with this scale invariant dynamics is
\be
\alpha= {1 \over 2} \log_2 F_{b=2} (x^*) \simeq 0.177352\ldots
\ee
that must be compared with the known exact value $\alpha=1/2$.

\begin{figure}
\centerline{\psfig{figure=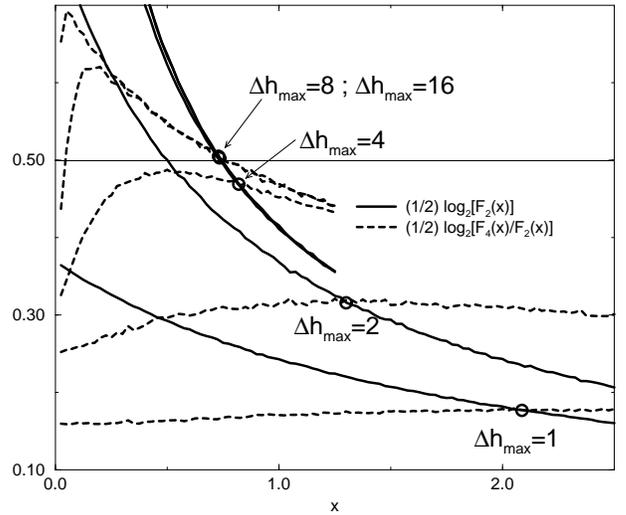,width=10cm,angle=-90}}
\caption{Results for $d=1$ and different $\Delta h_{max}$. The intersections
between solid (representing the $1/2 \log_2 F_2(x)$)
and dashed lines (representing the $1/2 \log_2 [F_4(x)/F_2(x)]$)
give $x^*$ on the horizontal axis and $\alpha$ on the vertical axis
(see Eq.~(\ref{alpha})). Observe that, as $\Delta h_{max}$ is increased, the
exact value $\alpha=1/2$ is rapidly approached.}
\label{Fig04}
\end{figure}

The apparent poor performance of the method is due to the assumption
that $\Delta h_{max}=1$,
which allows for full analytical treatment, but is clearly wrong.
The point is that even if at the microscopic level the dynamics is
of restricted type, the effective dynamics at generic scale defined
by the renormalization procedure will proliferate in a non restricted one.
Allowing larger steps ($\Delta h_{max}>1$) increases the number of
superficial configurations and makes the analytical determination
of the function $\omega^2(b,x_k)$ impossible.
Still this task can be performed numerically via simulation of systems
of such a small size.
Fig.~\ref{Fig04} reports the results obtained by considering increasing
values of $\Delta h_{max}$.
The value of $x^*(\Delta h_{max})$ converges already for $\Delta h_{max}=8$
to $x^* \simeq 0.726$,
corresponding to a value $\alpha=0.507\ldots$ in excellent agreement
with the exact value.
Further increases of $\Delta h_{max}$ do not change the results,
indicating that in the scale invariant dynamics the probability of steps
larger than 8 is negligible.

\subsubsection{$d>1$}

\begin{figure}
\centerline{\psfig{figure=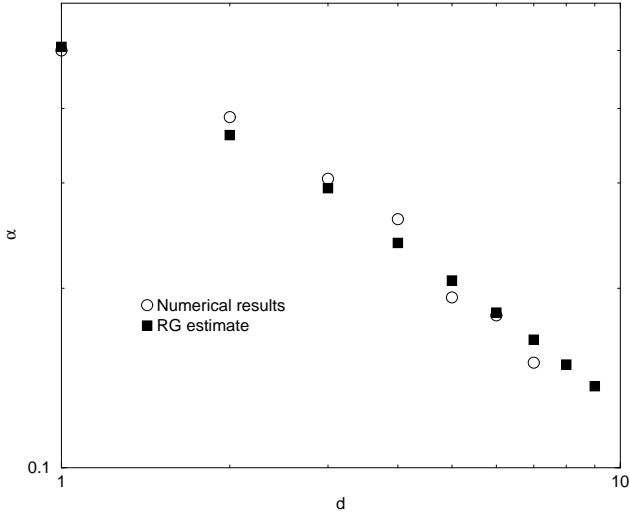,width=10cm,angle=-90}}
\caption{Value of $\alpha$ as a function of the dimension obtained by the
application of the method with small $b$, compared with numerical data by
Ala Nissila {\em et al}. The best fit to the RG values gives 
slope $-0.7$. We expect this exponent to converge to $-1$ for
large dimensions.}
\label{Fig05}
\end{figure}

The computation via Monte Carlo method of $\omega^2(2,x_k)$ and
$\omega^2(4,x_k)$ can be performed with very little computational effort
also in higher dimension.
We considered $d=1,\ldots,9$ with less than a week of CPU time of a
workstation.
The results are reported in Table I and summarized graphically in
Fig.~\ref{Fig05}.
We find a finite attractive fixed point for all dimensions, with an exponent
$\alpha$ in remarkably good agreement with the best numerical results
available\cite{debate}.
This is the first theoretical approach providing estimates for the roughness
exponent that match in all dimensions with numerics.
No anomalies are found for $d=4$ where other approaches find an
upper critical dimension.
The extrapolation to $d \to \infty$ suggests that the fixed point
is always stable and that $\alpha$ decreases with $d$ but remains always
nonvanishing.
The fixed point parameter $x^*$ grows exponentially with the dimension.
These results are confirmed by an analytical expansion of the method
in high dimensions, that is presented in subsection D.

\subsection{Robustness of the results}

\subsubsection{$b>2$}
In order to analyze the stability of the results upon increasing
the value of $b$ it is convenient to introduce the function
\be
\alpha_\ell(x) = {1 \over 2} \log_\ell F_\ell(x).
\label{alphal}
\ee
With this definition one can express the fixed point condition
Eq.~(\ref{FP})
as
\be
\alpha_b(x^*) = \alpha_{b^2}(x^*),
\label{FPl}
\ee
and see that the fixed point is stable if
\be
|R'(x^*)| = \left| 2 {\alpha'_{b^2}(x^*) \over
 {\alpha'_b(x^*)}}-1 \right| < 1,
\label{Ralpha}
\ee
i. e. 
\be
0<{\alpha'_{b^2}(x^*) \over \alpha'_b(x^*)} < 1.
\ee
Such a formula can be extended also to the case where the 
size of the larger system considered is not $b^2$ but a generic $b'>b$.
We study the stability of the results for growing $b$ by computing 
$\alpha_b(x)$ with $b_i=2,4,8,16,\ldots$ and imposing the consistency between
two successive $b_i$.
The value of $b$ indicated  in the plots is the smaller one;
for instance, $b=4$ label the results obtained imposing the consistency
between $b=4$ and $b'=8$.

\begin{figure}
\centerline{\psfig{figure=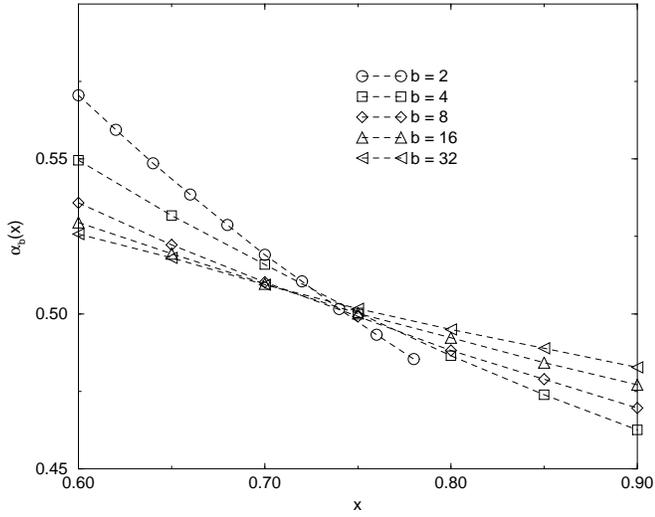,width=10cm,angle=-90}}
\caption{Behavior of the curves $\alpha_b(x)$ in $d=1$
(see Eq.~(\ref{alphal}) and (\ref{FPl})).}
\label{Fig06}
\end{figure}

\begin{figure}
\centerline{\psfig{figure=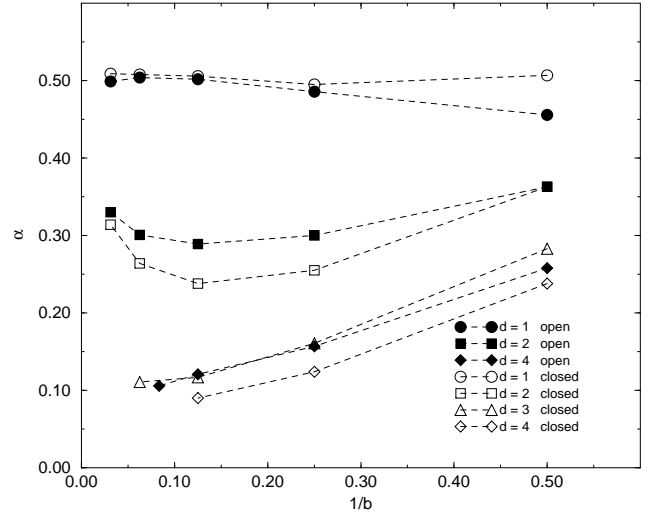,width=10cm,angle=-90}}
\caption{Plot of the RG estimate for the roughness exponent $\alpha$
as a function of the inverse system size. Empty symbols are for data obtained
with periodic boundary conditions. Full symbols refer to open boundaries.}
\label{Fig07}
\end{figure}

\begin{figure}
\centerline{\psfig{figure=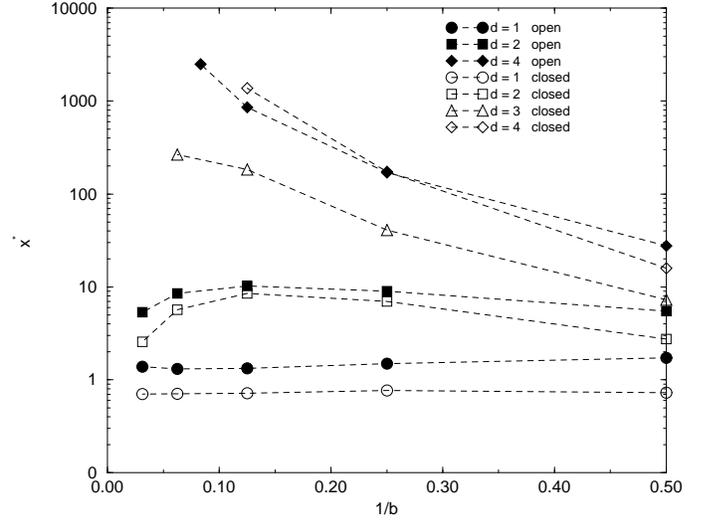,width=10cm,angle=-90}}
\caption{Value of the fixed point parameter $x^*$
as a function of the inverse system size. Empty symbols are for data obtained
with periodic boundary conditions. Full symbols refer to open boundaries.}
\label{Fig08}
\end{figure}

In Fig.~\ref{Fig06} we report the plot of the curves $\alpha_b(x)$ in $d=1$
for $b=2,4,8,16,32$.
Remarkably they all meet practically at the same point indicating that
$x^*$ and $\alpha$ virtually do not change by increasing the number of
cells.
Fig.~\ref{Fig07} reports the values of the exponent $\alpha$ in $d=1$
(empty circles). Observe that fluctuations are extremely small.
The value of the fixed point parameter $x^*$ is reported in Fig.~\ref{Fig08}
(empty circles).
Again it remains practically unchanged when $b$ grows.

In higher dimensions the results are less stable.
The values of $\alpha$ and of $x^*$ for $d=2, 3$ and 4 are reported in
Fig.~\ref{Fig07} and~\ref{Fig08}, respectively (empty symbols).
A clear trend is present for $d=2$: the exponent initially decreases as
$b$ is increased, then reaches a minimum and starts growing.
This behavior is reflected in the value of $x^*$ that first grows and then
decreases.

The decreasing part of the pattern is present in the analogous plots
for $d=3$ and $d=4$.
For large dimensions however, it is increasingly more time consuming
to perform the computation for large systems.
In particular for $d=4$, the largest system  that could be simulated is 
$b=16$ and for such a system size the trend is still decreasing.
Therefore it is not possible to decide from a numerical point of view
 whether for larger values of $b$ $\alpha$
would converge to zero or to a finite value.
These data do not provide any conclusive
 indication on whether $d=4$ is the upper critical
dimension for KPZ growth.
However, as it will be shown below, such a conclusion is ruled out
by the results with other parametrizations and by the analytical
large-$d$ expansion of the method.

The reason for the difference in the stability of the results for large
number of cells in $d=1$ and $d \geq 2$ is probably related to crossover
phenomena.
In the RG flow there are two competing fixed points; this reflects
the existence of two universality classes, strong coupling KPZ and EW.
In $d=1$ this fixed points are associated with the same roughness
exponent $\alpha=1/2$ and to similar values of $x^*$. Therefore,
any crossover phenomenon between the two fixed points has little effect
in our formalism.
In $d \geq 2$ instead, the two scale invariant dynamics are associated
with different exponents and also very different values of the parameter
$x^*$, which is finite for KPZ and infinite for EW.
We interpret the initial decrease in the value of $\alpha$ in the KPZ
case as the effect of a crossover
 caused by the presence of the EW fixed point.
It is not clear to us, however, why the fixed point found for $b=2$ is so
close to the results of the numerical simulations.

\subsubsection{Open boundary conditions}

The calculation of $\omega^2(b,k)$ can also be performed with open
boundary conditions, that is assuming that the height of the columns
outside the system which are in contact with the boundary is the same of
their neighbors inside the system. This means that no ``lateral growth''
event can be caused in the system by the environment around it.

The results are also presented in Fig.~\ref{Fig07}
and~\ref{Fig08} (filled symbols).
Interestingly, in this case the accuracy of the method for $b=2$
is not as good as for periodic boundary conditions, but the error remains
below 10\%, indicating a low sensitivity to the boundary conditions
even for small number of cells.
For larger number of cells the difference goes quickly to zero.

For higher dimensions the general dependence of $x^*$ and $\alpha$ on $b$
remains unchanged: In $d=2$ $\alpha$ is initially high, then decreases
and finally increases again.
The variations with $b$ are however less strong than when periodic 
boundary conditions are considered.
For $d=4$ it is more clear than in the case with periodic boundary conditions
that $\alpha$ does not converge to zero for large $b$.

\subsubsection{Other parametrizations of the dynamics}

\begin{figure}
\centerline{\psfig{figure=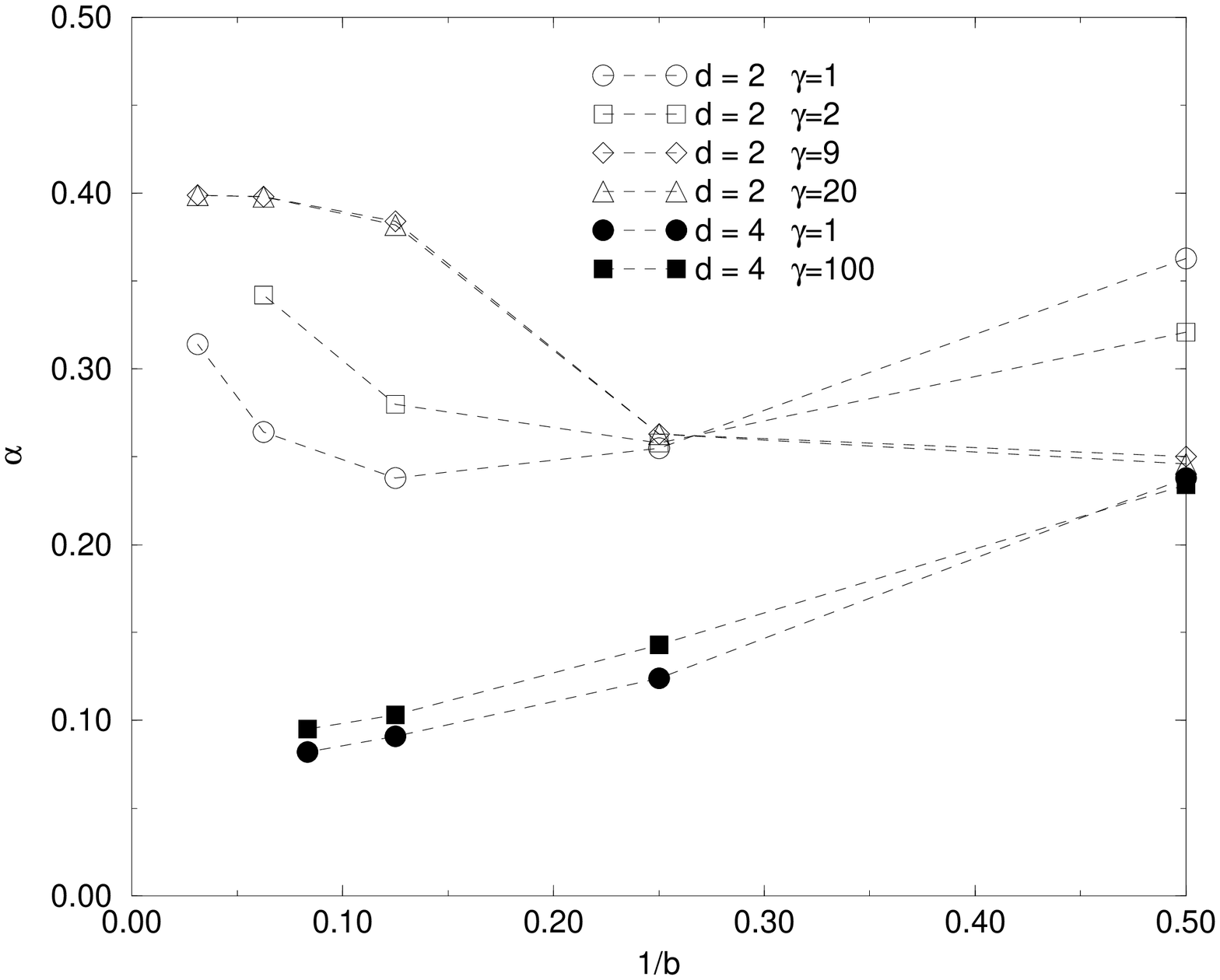,width=10cm,angle=-90}}
\caption{Value of the exponent $\alpha$ computed using the
parametrization Eq.~(\ref{gamma}) for the KPZ dynamics
for various values of $\gamma$ in $d=2$ and $d=4$.}
\label{Fig09}
\end{figure}

As stated above the parametrization~(\ref{r_i}) of the KPZ 
scale invariant dynamics
is by no means unique. Actually, given the problems of slow and
nonmonotonic convergence towards the asymptotic values, it turns out
clearly that the parametrization~(\ref{r_i}) is quite far from being
optimal and better parametrizations would help.
In order to keep things as simple as possible we started considering 
transition rates of the form
\be
r_i = 1 + x_k
\sum_{j n.n. i} \left\{\max\left[0,h(j)-h(i)\right]\right\}^{\gamma}
\label{gamma}
\ee
with $\gamma$ constant; for $\gamma=1$ it coincides with~(\ref{r_i}).
By comparing the values of $\alpha$ obtained with several $\gamma$
an interesting pattern can be spotted (Fig.~\ref{Fig09}).
While for small number of cells $b$ the estimate gets worse with increasing
$\gamma$, the opposite is true for large $b$.
For large values of $\gamma$ the estimate for $\alpha$ converges
quite rapidly.
For $\gamma=9$ and $\gamma=20$ we find on the largest systems $\alpha=0.399$,
suggestive of a convergence towards 0.4.
For $d=4$ the sizes that can be simulated are too small to allow
the determination of the asymptotic value of $\alpha$.
However, it is clearly seen that $\alpha$ does not go to zero as $b$ is
increased.

The same type of behavior is found by using an exponential parametrization
of the dynamics
\be
r_i = 1 + x_k \sum_{j n.n. i}
\exp\left\{\gamma \cdot \max\left[0,h(j)-h(i)\right]\right\}-1.
\label{exp}
\ee
In $d=2$ for large $\gamma$ the estimate of $\alpha$ on the largest
system is 0.399, exactly as with Eq.~(\ref{gamma}).
In $d=4$ again we cannot precisely determine where $\alpha$ is converging
to.
Again the data strongly suggest that this limit is finite.

The study of these two alternative parametrizations of the dynamics
consistently indicates a value of $\alpha=0.399$ in $d=2$ and a finite
$\alpha>0$ in $d=4$ suggesting strongly that 4 is not the upper critical
dimension of the KPZ.

One could in principle imagine a parametrization of the effective
dynamics, more in the spirit of the KPZ equation, of the type
\be
r_1 = 1 + \nu_k |\nabla^2 h(i)| + \lambda_k [\nabla h(i)]^2
\label{r_i2}
\ee
However, there is no reason for believing that such a parametrization
would be better for the KPZ rough phase; additional operators
are very likely to be present in the scale invariant dynamics.
Moreover the dynamics described by Eq.~(\ref{r_i2}) is plagued by
numerical instabilities, as pointed out by Bray and Newman\cite{Bray97}.

\subsection{The $d\to\infty$ limit and the upper critical dimension.}

The results presented so far show that $\alpha>0$ even for large
number of cells in $d=4$,
thus indicating that 4 is not the upper critical dimension
for KPZ growth processes.
By using the RG procedure it is actually possible to go beyond this
numerical conclusion:
{\it The existence of any finite upper critical dimension can be ruled out.}
This result is obtained when the function $\omega^2(b,x_k)$ is computed
analytically in the large-$d$ limit.
The basic fact allowing this calculation is that when $d \gg 1$ one 
expects $\alpha \ll 1$, which suggests that surface fluctuations 
are small
\be
\omega(b,x_k) \sim b^\alpha \simeq 1+\alpha \ln b + O(\alpha^2).
\ee
For small $b$ one may reasonably account for the fluctuations
of the interface by considering only two possible values of $h(i)$,
$h_0$ (``low sites'') and $h_0+1$ (``high sites'') (Fig.~\ref{Fig11}).
Starting from a flat surface ($h(i)=0, \forall i$), one considers
growth events occurring 
according to the rates Eq.~(\ref{r_i}), with the restriction
that no block can be deposited on top of an already grown one.
Only when the whole layer at height 1 is grown one allows growth to level 2
and so on.
This approximation allows the analytical evaluation
of $\omega^2(b,k)$, the identification of the fixed points and the study
of their stability.
We will check a posteriori the consistence of the results with the assumption,
and see that the existence of a finite upper critical dimension can be
excluded.
Let us now present the details of the calculation.

Within the ``two layers'' approximation it is convenient to group
together all configurations with the same number of high sites:
we will call ``state'' $n$ the set of all surface configurations
with $n$ sites at height $h_0+1$ and the remaining $b^d-n$ sites at height
$h_0$.
The state $n=0$, corresponding to a flat surface,
is equivalent to the state with $n=b^d$.
This classification is useful because the only transitions permitted
from state $n$ are those to state $n+1$.
The master equation for the probability $\rho_n$ of being in state $n$
(i.e. of having any of the configurations with $n$ high sites)
is then greatly simplified
\be
\partial_t \rho_n = \rho_{n-1} r(n-1 \to n) - \rho_n r(n\to n+1).
\label{ME}
\ee
$r(n \to n+1)$ is the average of all the rates~(\ref{r_i}) for the
growth processes that transform one configuration with $n$ high sites in
one with $n+1$ of them.
We can write this quantity in the form
\be
r(n \to n+1)  = (b^d-n) + x_k \Omega_n.
\ee
The first term on the right hand side is simply the total 
rate of vertical growth [1 in Eq.~(\ref{r_i})] for configurations
with $n$ high sites. Observe that it
 is obviously equal to the number ($b^d-n$) of sites
where vertical growth is allowed.
$x_k \Omega_n$ is the rate for lateral growth: $\Omega_n$ is the average
number of lateral walls in configurations with $n$ high sites.
Its precise computation is not easy, since it would require the knowledge
of the stationary probability for each configuration belonging to state $n$.
However, when $x_k=0$ the computation is trivial since growth occurs
only via uncorrelated deposition and high and low sites are randomly
distributed.
The number of low sites, where growth is allowed, is $b^d-n$;
each of them has $2d$ neighbors which are occupied with probability $n/b^d$.
Hence the average number of lateral walls is
\be
\Omega_n = (b^d-n) 2d {n \over b^d}
\label{Omega_n}
\ee
The form of $\Omega_n$ for $x\neq 0$ is in general more complicated,
but a numerical computation  for large dimensions,
namely for $d=7$, shows that Eq.~(\ref{Omega_n})
 is a good approximation for all values of $x_k$ (Fig.~\ref{Fig12}).
We assume the validity of Eq.~(\ref{Omega_n}) for all values
of $x_k$. This leads to
\be
r(n \to n+1) = (b^d-n) \left[1 + 2d {n \over b^d} x_k \right].
\ee
The stationary solution of Eq.~(\ref{ME}) is
\be
\rho_n = \rho_{0} \frac{r(0 \rightarrow 1)} {r(n \rightarrow n+ 1)},
~~~~n=1,\ldots,b^d-1.
\label{ro}
\ee
By imposing the normalization condition $\sum_{n=0}^{b^d-1} \rho_n =1$
and approximating the sum by an integral one obtains
\be
\rho_0 = \left\{  1 + \frac{b^d}{2 d x_k} \left[ 
2 d \ln b + \ln  \left(\frac{1+2dx_k}{b^d+2 d x_k} 
\right) \right]
          \right\}^{-1}.
\label{r0}
\ee
Equations (\ref{ro}) and (\ref{r0}) provide a complete description 
of the stationary probability density.
Given that the roughness of all configurations with
$n$ high sites is $n/b^d (1-n/b^d)$,
the total roughness of the surface can be computed as
\be 
\omega^2(b,x_k) = \sum_{n=0}^{b^d-1} \rho_n
\left(1-{n \over b^d}\right) {n \over b^d}. 
\ee
Using the fact that $b^d \gg 1$ and assuming
$d x_k \gg 1$, we obtain
\be
\omega^2(b,x_k)  = \rho_0 {b^d \over 2 d x_k}.
\label{omega2}
\ee
Inserting Eq.~(\ref{omega2}) with $b=2$ and $b=4$
in the fixed point equation~(\ref{FP})
yields, to leading order in $d$,
\be
x^* = 2^{d+1} \ln 2.
\label{x*d}
\ee
The assumption $d x_k \gg 1$ is therefore self-consistent for sufficiently
large $d$.
Notice that an exponential dependence of $x^*$ on $d$ was already found in
the numerical implementation of the method (Fig.~\ref{Fig08}).
Using Eq.~(\ref{alpha}) one obtains the value of the roughness exponent
\be
\alpha \simeq {1 \over 3(\ln 2)^2} {1 \over d}.
\label{alphad}
\ee
Finally, by computing the derivative of the RG transformation at the
fixed point
\be
R'(x^*) = -1 + {1 \over 2 \ln 2} {1 \over d} + O(1/d^2),
\ee
we see that the fixed point is attractive for all finite dimensions.

In conclusion we find that for large-$d$ the RG has a fixed point
$x^*$ corresponding to an exponent $\alpha \sim 1/d$ and therefore
strictly positive in all finite dimensions.
On the contrary the existence of a finite upper critical dimension
would have implied, for $d>d_c$, either the absence of a finite fixed point
or its instability.

At this point we must use the analytical result to check the consistency
of the two layers assumption. Such assumption is correct provided the
rate of its violation is negligible for all values of $n$.
Processes violating the assumption are those in which an event of vertical
growth occurs on top of an high site.
Their rate in state $n$ is $r_{up}=n$,
that must be compared to the total rate of processes not violating
the restriction $r(n \to n+1)$ computed for $x=x^*$.
By imposing $r_{up}(n) \ll r(n \to n+1)$ we get
\be
n \ll (b^d-n) \left(1+{2dn \over b^d} x^*\right).
\ee
Let us consider $n=b^d-1$ which is the situation that maximizes $r_{up}$
and minimizes $r(n \to n+1)$.
Then
\be
b^d-1 \ll 1+{2d (b^d-1) \over b^d} x^*.
\ee
Since $b^d \gg 1$, this means
\be
b^d \ll 2dx^* \sim 2^{d+2} d.
\ee
Hence the two layers assumption is correct for $b=2$ but fails for $b=4$.
Therefore the value of $\omega^2(4,x_k)$ is systematically
underestimated by Eq.~(\ref{omega2}) since fluctuations involving more than
two layers are neglected despite being likely.
The consequence of this on our results is understood by considering
Eq.~(\ref{RGeq}).
In such a formula we estimate correctly the left hand side, while the
right hand side is underestimated (Fig.~\ref{Fig13}).
Since the fixed point parameter $x^*$ and the exponent $\alpha$ are given
by the intersection of the curves it is clear that we get
an upper bound  for $x^*$ and a lower bound for $\alpha$.
This is confirmed by the comparison of our estimate of $\alpha$,
Eq. (\ref{alphad}),
and $x^*$ with the numerical results of Ala-Nissila {\em et al.}
and the value of $x^*$ computed numerically for $d=1,\ldots,9$
(Fig.~\ref{Fig14}).

These results have been obtained for the smallest values of $b$,
namely $b=2$.
In the previous sections we showed that for low dimensions the results
for small $b$ are in good agreement with numerics, but for larger $b$
there are deviations.
A very reasonable question therefore concerns the robustness of the
large-$d$ results when $b$ grows.
As we have shown above the two layers approximation breaks down for $b>2$.
In order to extend the above calculation to larger $b$ one should replace
the two layers approximation with some less restrictive but still
doable calculational scheme.
We have not been able to fulfill this task and hence we cannot
directly show whether a fixed point exists for finite $x^*$ when $b \to
\infty$.
However, the two layers assumption is valid for any $b$ in the neighborhood
of the fixed point  $x^*=\infty$ that gives a flat
surface $\alpha=0$. This fixed point exists in any dimension, and
its stability can be safely analyzed using 
the previous two layer assumption as follows.

Let us introduce $\epsilon=1/x$. The derivative of the RG transformation
at the fixed point $\epsilon^*=0$ is (see Eq.~(\ref{Ralpha}))
\be
R'(\epsilon=0)= {2 \alpha'_b(\epsilon=0) \over \alpha'_{b^2}(\epsilon=0)}-1
\ee
where now prime indicates derivative with respect to $\epsilon$.
To first order in $\epsilon$ we have (see Appendix II)
\be
\alpha_{b^2}(\epsilon) = {1 \over 4 \ln b} \ln \left[
1 + c \mu \epsilon b^{2d+2} \right].
\label{eq54}
\ee
Then
\be
\alpha'_{b^2}(\epsilon=0) = {1 \over 4 \ln b} c \mu b^{2d+2}.
\ee
Analogously
\be
\alpha'_b(\epsilon=0) = {1 \over 2 \ln b} c \mu b^{d+1}.
\ee
Hence
\be
R'(\epsilon=0) = b^{d+1}-1 \gg 1
\ee
and the fixed point corresponding to $\alpha=0$ is unstable.
As a consequence, a finite fixed point with $\alpha>0$ must exist and be
stable when $b \to \infty$ for any large and finite $d$.
This supports the conclusion that there is no finite upper critical
dimension.

\section{RG for the Edwards-Wilkinson dynamics}

So far we have applied the new RG method to a KPZ-like
dynamics.
Now we intend to show that it is more general and can be applied 
to growth mechanisms belonging to universality classes other 
than KPZ. In particular we study in this section its application
to the exactly solvable, Edwards-Wilkinson (EW) equation 
for which the roughness exponent is known in any dimension.  
In particular $\alpha=1/2$ in $d=1$, while $\alpha=0$ for
$d \geq 2$ with logarithmic corrections at $d=2$.
        
The parametrization Eq.~(\ref{r_i}) of the dynamics describes a growth model
where only deposition events can take place and the symmetry between up
and down in the $h$ direction is clearly broken.
Such a dynamics is inherently out of equilibrium and therefore
cannot accommodate the scale invariant dynamics of the Edwards-Wilkinson
growth process, which is an equilibrium one, with growth rules symmetric
along the growth direction.
We now introduce a generalized dynamics which
admits the KPZ and EW dynamics as particular limiting cases.

Let us consider the quantities
\begin{eqnarray}
K_d(i) & = & \sum_{j nn i} \max[0,h(j)-h(i)] \nonumber \\
K_u(i) & = & \sum_{j nn i} \max[0,h(i)-h(j)].
\end{eqnarray}
In the KPZ case described so far we have allowed
only deposition of particles and written
\be
r_i = 1+x_k K_u(i).
\ee

We now allow also for evaporation of particles. 
That is, we consider the transition rate for site $i$ as
\be
r_i = 1+x_k |\epsilon K_u(i)-(1-\epsilon) K_d(i)|
\ee
and with probability
\be
P_b=1/r_i
\ee
a random deposition/evaporation event takes place ($h_i \to h_i +1$ with
probability $\epsilon$ and $h_i \to h_i-1$ with probability $1-\epsilon$),
while with probability
\be
P_l=x_k |\epsilon K_u(i)-(1-\epsilon) K_d(i)|/r_i
\ee
we have a ``lateral'' event
\be
h_i \to h_i +{\epsilon K_u(i)-(1-\epsilon) K_d(i) \over
|\epsilon K_u(i)-(1-\epsilon) K_d(i)|}.
\ee
For $\epsilon=1$ only deposition is allowed and we have the transition
rates for the KPZ dynamics.
For $\epsilon=1/2$, we have up-down (deposition-evaporation) symmetry
and the rates are 
\be
r_i = 1+x_k |\nabla^2 h(i)|
\ee
where $\nabla^2 h(i)=[K_u(i)-K_d(i)]/2$ is the discretized Laplacian
evaluated at site $i$. Therefore we expect this case to correspond to
EW dynamics.
Let us verify that for $\epsilon=1/2$ the average interface velocity
does not depend on the interface
configuration (which is a basic property of EW dynamics)
\be
v = {1 \over  L^d} \sum_i v_i = {1 \over  L^d} \sum_i r_i \Delta h_i.
\ee
Since
\begin{eqnarray}
\Delta h_i & = & P_b \cdot 0 + P_l \cdot {K_u(i)-K_d(i)\over |K_u(i)-K_d(i)|}
\nonumber \\ & = &
{1 \over r_i} \left(x |K_u(i)-K_d(i)| { K_u(i)-K_d(i)\over
|K_u(i)-K_d(i)|} \right) \nonumber \\
& = & 
{1 \over r_i} \left(x [K_u(i)-K_d(i)] \right)
\end{eqnarray}
and as  $\sum_i K_u(i) = \sum_i K_d(i)$, we have that
\be
v = {x \over  L^d} \sum_i [K_u(i)-K_d(i)] = 0.
\ee

With this generic dynamics we can perform the RG procedure exactly as in
the KPZ case. The evaluation of the function $\omega^2(b,x)$ is carried out
again using small Monte Carlo simulations with periodic boundary conditions.
The results are reported in Fig.~\ref{Fig15}.
For $d=1$ the value of $\alpha$ for $b=2$ is $\simeq 0.4$ below the exact
value $1/2$, but for $b>2$ the correct value is rapidly approached.
The situation is completely different  in $d=2$.
In such case for $b=2$ the exponent $\alpha$ is around 0.25, but when
$b$ is increased, the fixed point is shifted monotonically towards
$\infty$.
In $d=3$ the behavior of the fixed point for finite $x^*$ is similar.
From these plots we can conclude that the behavior of the EW dynamics
is very different in $d=1$ and $d>1$.
For $d=1$ there is a stable fixed point with $\alpha=1/2$.
For $d>1$ there is no stable fixed point with $\alpha \neq 0$.
Hence the RG method is able to capture the difference between KPZ and
EW dynamics and correctly describes both the rough and the flat phases
of the Edwards-Wilkinson growth.
We speculate that the reason why one needs to consider $b>2$ is probably
related to the fact that on a system of size $b=2$ the discretized
Laplacian and square gradient take the same form.

\section{Discussion} 
In the previous sections we have introduced a general method for
studying surface growth models by means of a real space renormalization
group procedure.
The anisotropy of the scale invariant properties of surfaces
makes the very definition of the RSRG highly nontrivial, since
the direct integration of degrees of freedom at small scale cannot be
performed.
For this reason we had to devise an alternative route: the main 
ingredient is that the integration of degrees of freedom is performed
implicitly by imposing the self-consistency between two descriptions
of the same object at different levels of coarse-graining.
The application of such an approach to the KPZ dynamics yields several
results that can be summarized as follows.
\begin{itemize}
\item
The scale invariant dynamics is identified and parametrized as
a function of the ``lateral growth'' parameter $x$. 
This parameter turns out to have a non-trivial attractive fixed point
under RG transformation for all dimensions.
\item The KPZ roughness exponent $\alpha$ estimated for small $b$ is in
very good agreement with large scale simulations of discrete models.
\item
For larger values of $b$ the estimate of $\alpha$ is stable in $d=1$ while it
changes noticeably for $d \geq 2$, presumably owing to crossover effects.
When $b \to \infty$ it converges  towards the correct result.
\item
The results are robust with respect to changes in the parametrization
of the dynamics and in the boundary conditions.

\item
No evidence is found of the existence of an upper critical dimension 
for KPZ. Moreover, we show very strong evidence that no such
an upper critical dimension exists.

\item
By changing the nature of the parametrization of the dynamics at generic
scale, the method is able to describe the EW dynamics and capture the
existence for it of an upper critical dimension above which only 
a trivial (flat) phase exists.

\end{itemize}

Regarding the general nature of the approach it is worth remarking that
the key point in the method is the identification of the scale
invariant dynamics.
In some sense the procedure can be seen as a kind of finite size scaling
approach allowing for the evaluation of scaling exponents via the
extrapolation of small size MC simulations.
However the crucial point is that the MC data do not directly determine 
the exponent; they rather allow the identification of the scale invariant
parameters of the dynamics which in turn determines the exponent.

With respect to the estimates of the roughness exponent for small $b$,
it is remarkable that the accuracy (in the sense of the discrepancy with
known numerical results) seems to be the same in all dimensions $d \geq 2$.
This is not the typical situation in ordinary critical phenomena, where
usually RSRG methods fail in high dimensions.
There are at least two reasons why usual RSRG schemes are inaccurate
in high dimensions:
 
a) The necessity of defining an explicit geometrical mapping
between degrees of freedom at two scales (spanning rule, majority
rule, bond-moving, etc.).
 
b) The presence of relevant fields and the need to 
compute exponents from the derivatives of 
the RG transformation at the fixed point.

Due to the success of field theory in high dimensions 
(close to $d_c$) in usual critical phenomena, RSRG methods have
been mostly devised to work in low dimensions where the
predictions of the $\epsilon$-expansion become less reliable. 
RSRG methods based on an explicit geometric mapping (block-spin
transformation) are quite accurate (and sometimes even 
exact) close to the lower critical dimension. 
Problems related with this geometric transformations
become worse and worse as the dimension increases. 
In our perspective, the only limitation has to do with the quality of 
the parametrization of the RG transformation. 
For example, the parametrization of the RSRG transformation for
ferromagnetic systems based on the Migdal transformation 
of Ising spins gives inaccurate results in high $d$. 
However, if one uses the parametrization of the $\phi^4$ 
theory, one recovers $d_c=4$ within the RSRG Migdal approach 
\cite{LM} even for ferromagnetic systems.
 
In any case, notice that our RSRG method does not need
an explicit geometrical definition of the RG transformation.
Therefore it bypasses the problems related to a). 
In some sense, this is similar to the phenomenological RG method
\cite{DV}
where the RG transformation is defined implicitly through finite
size scaling arguments. Remarkably, phenomenological RG calculations
are quite accurate.

With respect to point b), as discussed above 
the absence of relevant fields makes truncation errors 
much less important than in ordinary applications of the RG.
Furthermore we note that in the KPZ problem one has to compute
exponents depending only on the RG transformation at
the fixed point, i. e. the critical parameter.
This is profoundly different from what happens in Ising-like problems,
where some exponents (as, for example, the correlation length exponent
$\nu$) depend on the derivatives of the RG transformation around it:
As a consequence $\nu$-type exponents are rather difficult to estimate
since, even if the location of the fixed point is determined accurately,
the computation of the derivatives is much less precise.
No exponents of such type exist in the KPZ case.
This is, in our opinion, a further reason for the great accuracy of the new
method with respect to the usual RSRG. 

As a final point, it is worth discussing the current limitations of the
method.
It is clear from the results presented, that a most important role in the
method is played by the choice of the parametrization of the scale invariant
dynamics.
This is particularly true since one deals with a monoparametric description
of the growth process: If one could easily introduce several parameters
and study their flow under the RG, the stability of the results when details
are changed would be greatly improved.
Within the present framework, the inclusion of additional parameters is
however not straightforward.
The problem is that additional RG equations are provided only by the
use of equation~(\ref{main}) with $b$, $b^2$, $b^3$ and so on.
This requires the computation of $\omega^2(b,k)$ on systems whose size
becomes quickly prohibitively large.
A remarkable improvement of the method would therefore be the identification
of additional RG transformations independent from Eq.~(\ref{main}).
Despite this difficulty, we believe that the theoretical framework
presented here constitutes an important new element in the field of
surface growth and deserves further investigation, in particular with
respect to possible applications to other open problems,
and generalization to deal with time-dependent properties.

In summary, in this paper we have presented a real space
renormalization group method developed to deal with surface growth
processes. The new method overcomes the difficulties inherent 
to standard real space renormalization group analysis of anisotropic
situations. It is based on the definition of anisotropic blocks
of generic scale and of a parametrized effective dynamics for the
evolution of such blocks. Imposing the surface width to be the same when 
using different scales (different block sizes) we write a renormalization
group equation. Its associated fixed points define the scale
invariant effective dynamics and permit to determine the roughness
exponent $\alpha$.

We have employed the new method to study the Kardar-Parisi-Zhang   
and the Edwards-Wilkinson
universality classes. In particular, for KPZ we compute the
$\alpha$ exponent in dimensions from $d=1$ to $d=9$. The results
are in very good agreement with the best numerical estimates
in all dimensions.
Moreover, we present analytical calculation excluding 
the possibility of KPZ having a finite upper critical dimension.
On the other hand, well known results for the EW universality class
are obtained, confirming  the generality of the method. 

\section{Acknowledgments}

We acknowledge interesting discussions with A. Gabrielli, A. Maritan,
G. Parisi, A. Stella, C. Tebaldi, G. Bianconi, and A. Vespignani.
This work has been partially supported by the European network
contract FMRXCT980183, and by a M. Curie fellowship,
ERBFMBICT960925, to M. A. M.

\appendix

\section{The RSRG method for large $b$}
In this Appendix we investigate the behavior of the method for
large values of $b$.
At the fixed point $x^*$, Eq.~(\ref{RGeq}) reads
\be
\omega^2(b^2,x^*)=c\omega^4(b,x^*)+2\omega^2(b,x^*).
\label{o1}
\ee
If we now assume that, for $b\gg 1$, 
\be
\omega^2(b,x)\simeq b^{2\alpha}\left[A(x)+B(x)b^{-\omega}+
\ldots\right]
\ee
as it should, we find that Eq. (\ref{o1}) becomes
\begin{eqnarray}
&&A(x^*)[1-cA(x^*)]-2cA(x^*)B(x^*)b^{-\omega} \nonumber \\
&&-2A(x^*)b^{-2\alpha}+
{\rm subleading~terms}=0
 \end{eqnarray}
We see that for $b\to \infty$ the fixed point tends to
\be
x^*_\infty:~~~cA(x^*_\infty)=1
\label{xstar}
\ee
whereas for large but finite $b$ 
\be
x^*_b=x^*_\infty-\frac{2D(x^*_\infty)}{A'(x^*_\infty)}b^{-\Delta}
\ee
with 
\begin{eqnarray}
\Delta &=&\min\{\omega,2\alpha\}~~\hbox{and}~~
D(x^*) \nonumber \\
&=&\left\{\begin{array}{cc}
B(x^*) & \hbox{ if $\omega< 2\alpha$,}\\
A(x^*)+B(x^*) & \hbox{ if $\omega= 2\alpha$,}\\
A(x^*) & \hbox{ if $\omega> 2\alpha$.}\end{array}\right.
 \end{eqnarray}
The RG estimate $\hat\alpha$ of the exponent $\alpha$ is given 
by
\begin{eqnarray}
&&\hat\alpha=\frac{\ln F_b(x^*_b)}{2\ln b}=
\nonumber \\
&&\alpha+\frac{\ln
\left[b^{-2\alpha} + cA(x^*_b)+cB(x^*_b)b^{-\omega}+\ldots\right]}{2\ln b}=
\nonumber \\
&&\alpha+\frac{\ln \left[1+b^{-2\alpha}
-2cD(x^*_\infty)b^{-\Delta}+ cB(x^*_\infty) b^{-\omega}+
\ldots\right]}{2\ln b}
\end{eqnarray}
and converges to the exact value for $b\to\infty$.
Note that only the finite $b$ corrections depend on $x^*$.
In particular, if $\omega<2 \alpha$
\be
\hat\alpha=\alpha-{c B(x^*_\infty) b^{-\omega} \over 2 \ln b}
\label{20}
\ee
If $\omega>2 \alpha$
\be
\hat\alpha=\alpha-{b^{-2 \alpha} \over 2 \ln b}
\ee
With respect to the stability of the fixed point one finds
\be
R'(x^*)=-b^{-2\alpha}-\left(\frac{B}{A}+\frac{B'}{A'}\right)
b^{-\omega}+\ldots
\label{22}
\ee
This means that, as it should be expected, the fixed point becomes
more and more stable as $b$ increases: for larger $b$ fewer 
RG iterations are necessary to reach the scale invariant regime.

These results for large $b$ are not surprising.
When the systems become large, the effect of the boundary conditions
is clearly small and also the choice of the parametrization of the
scale invariant dynamics tends to become irrelevant, since parameters
that are not included in the explicit parametrization are generated
by the RG procedure on large sytems.
The formulae~(\ref{20}-\ref{22}) certify that the RG method
is asymptotically correct.

\section{Computation of $\omega^2(b,\epsilon_k)$.}

In this Appendix we present the derivation of Eq.~(\ref{eq54}).
Let us consider $b$ and $d$ arbitrarily large but finite, so that
$\epsilon_k \ll b^d$.
We have
\be
\sum_{n=1}^{b^d-1} \rho_n = \rho_0 b^d \sum_{n=1}^{b^d-1}
{1 \over \Omega_n/\epsilon_k +b^d-n} = \rho_0 b^d \epsilon_k g +
{\cal O}(\epsilon_k^2),
\ee
where
\be
g=\sum_{n=1}^{b^d-1} {1 \over \Omega_n}.
\ee
Hence
\be
\rho_0 = {1 \over 1 +b^d \epsilon_k g} +{\cal O}(\epsilon_k^2)
\ee
and
\be
\rho_n = {b^d \epsilon_k \over \Omega_n} + {\cal O}(\epsilon_k^2). 
\ee
The roughness of a system of size $b$ is therefore
\begin{eqnarray}
\omega^2(b,\epsilon_k) & = &
\sum_{n=1}^{b^d-1} \rho_n {n \over b^d}
\left(1 - {n \over b^d} \right) \nonumber \\
& \approx & b^d \epsilon_k \int_{1/b^d}^{1-1/b^d}
{y(1-y) \over \Omega_{yb^d}} dy.
\end{eqnarray}
Fog large $b$ and $d$ and infinitely strong lateral growth parameter
$1/\epsilon_k$, the set of high sites will form, when $n \to 0$ a
$d$-dimensional hypersphere.
Hence $\Omega_n$ will scale as the perimeter of such hypersphere
\be
\Omega_n \sim n^{(d-1)/d}
\hspace{2cm}
n \to 0.
\ee
Similarly for $n \to b^d$ the low sites will form a shrinking hypersphere
and $\Omega_n \sim (b^d-n)^{(d-1)/d}$ for $n \to b^d$.
Hence it is reasonable to assume
\be
\Omega_{y b^d} = b^{d-1} {\hat \Omega}(y)
\ee
with ${\hat \Omega}(y) \sim y^{(d-1)/d}$ for $y \to 0$
and ${\hat \Omega}(y) \sim (1-y)^{(d-1)/d}$ for $y \to 1$.
The form of ${\hat \Omega}(y)$ for intermediate values of $y$ is not
known, but we expect it to be nonsingular and not dependent on $b$.
In conclusion
\be
\omega^2(b,\epsilon_k)  = b^{d+1} \epsilon_k \mu
\ee
with
\be
\mu = \int_0^1 dy {y (1-y) \over {\hat \Omega}(y)}
\ee
a finite geometrical constant.

\begin{figure}
\centerline{\psfig{figure=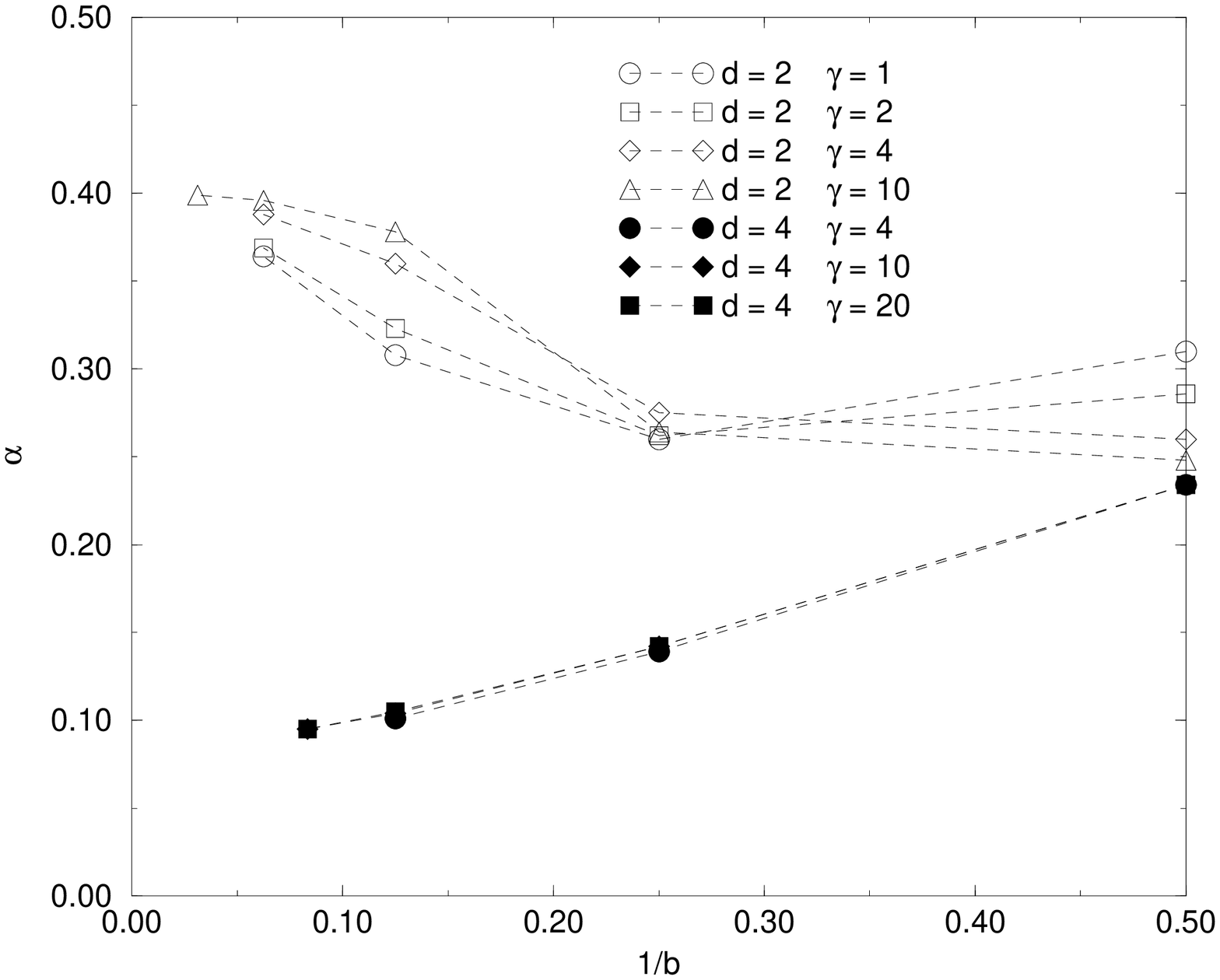,width=10cm,angle=-90}}
\caption{Value of the exponent $\alpha$ computed using the
parametrization Eq.~(\ref{exp}) for KPZ dynamics
for various values of $\gamma$ in $d=2$ and $d=4$.}
\label{Fig10}
\end{figure}

\begin{figure}
\centerline{\psfig{figure=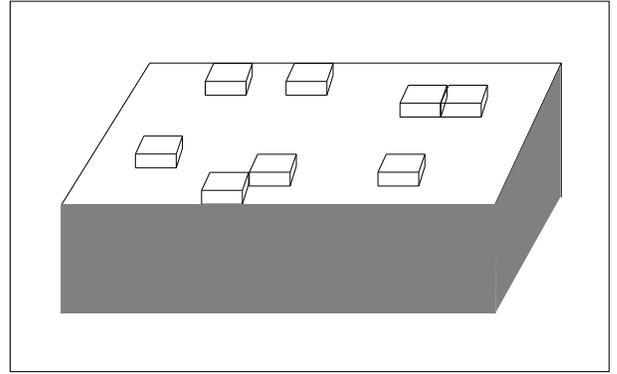,width=8cm,angle=-90}}
\caption{Representation of a surface configuration with 8 ``high'' sites
and $b^d-8$ ``low''sites.}
\label{Fig11}
\end{figure}

\begin{figure}
\centerline{\psfig{figure=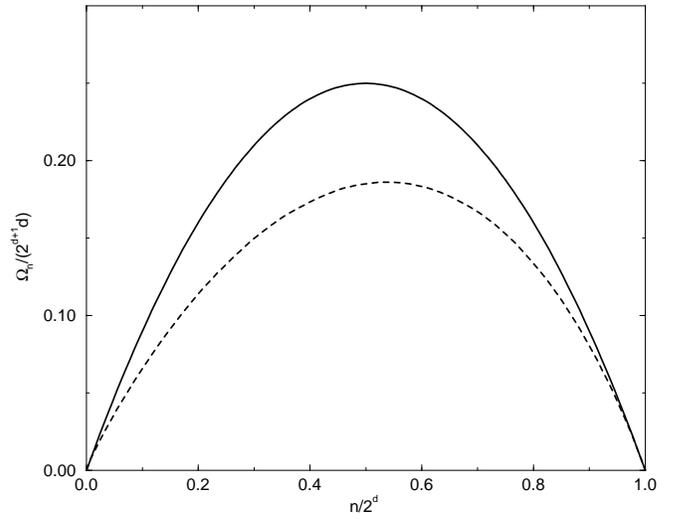,width=10cm,angle=-90}}
\caption{Comparison of the form of the function $\Omega_n$ computed with
$x=0$, used in the analytical calculation (solid line),
with the function determined numerically in $d=7$ for $x=\infty$ (dashed
line).
For intermediate values of $x$, the function
$\Omega_n$ lies between the two curves plotted.}
\label{Fig12}
\end{figure}

\begin{figure}
\centerline{\psfig{figure=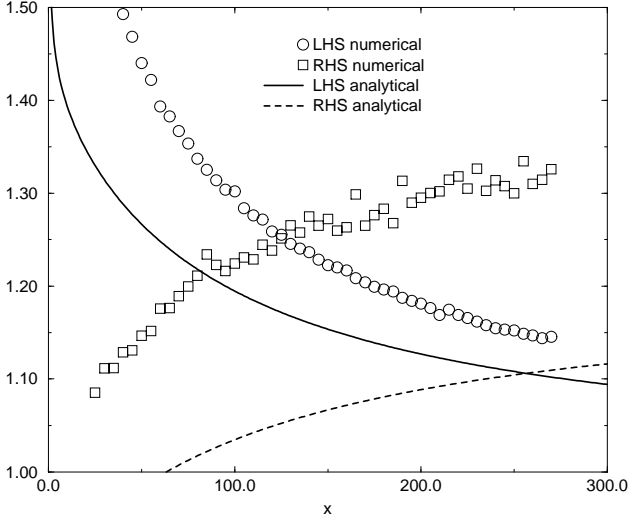,width=10cm,angle=-90}}
\caption{Plot of the left and right hand sides of Eq.~(\ref{RGeq}) as
computed numerically in $d=7$ and analytically (Eq.~(\ref{omega2})).}
\label{Fig13}
\end{figure}

\begin{figure}
\centerline{\psfig{figure=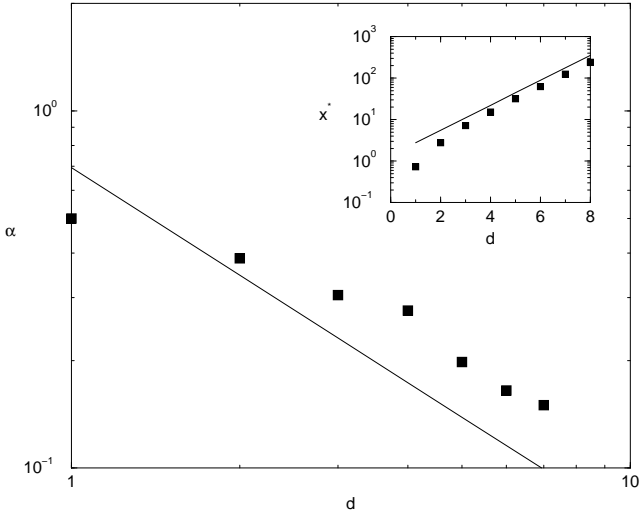,width=10cm,angle=-90}}
\caption{Comparison of the dependence of $\alpha$ on $d$ computed with 
the RG in the large-$d$ limit Eq.~(\ref{alphad})
with the numerical data of Ala-Nissila {\em et al.}. In the inset, plot of
the value of the fixed point parameter $x^*$ computed numerically for
small $d$ and compared with the analytical expression Eq.~(\ref{x*d}).}
\label{Fig14}
\end{figure}

\begin{figure}
\centerline{\psfig{figure=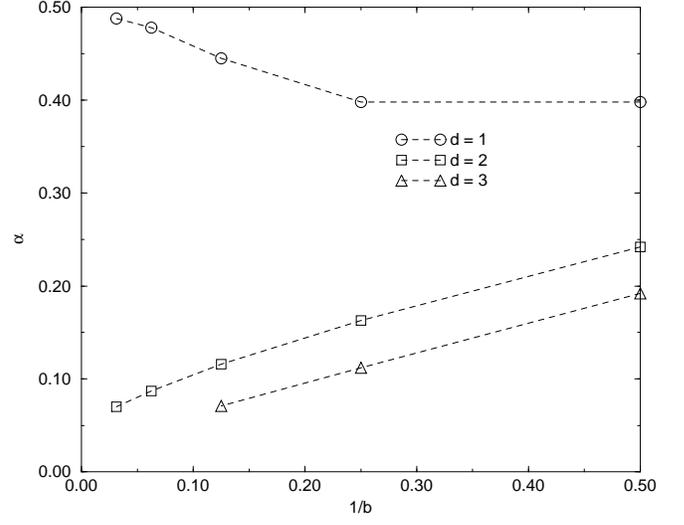,width=10cm,angle=-90}}
\caption{Plot of the RG estimate for the roughness exponent $\alpha$
as a function of the inverse system size for the Edwards-Wilkinson dynamics.}
\label{Fig15}
\end{figure}

\begin{table}
\begin{center}
\begin{tabular}{c|ccccccccc}
$d$                &1    &2    &3    & 4   & 5   &6    &7    &8    &9\\ \hline 
$x^*$              &0.726&2.77 &6.96 &15.91&31.96&63.5 &124.5&242  &468\\
$\alpha_{\rm RG}$  &0.507&0.363&0.294&0.238&0.206&0.182&0.164&0.149&0.137\\
$\alpha_{\rm num}$ &0.5  &0.387&0.305&0.261&0.193&0.18 &0.15 & -   & -  \\
\end{tabular}
\caption{Values of the fixed point parameter $x^*$ and of the roughness
exponent computed with the RG with $b=2$, compared with the numerical
results of Ala-Nissila {\em et al.}}
\end{center}
\label{table}
\end{table}


\end{document}